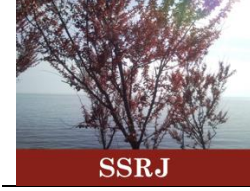

**The Mediating Effect of Blockchain Technology on the Cryptocurrency Purchase Intention**

İbrahim Halil Efendioğlu
Gaziantep Üniversitesi
efendioglu@gantep.edu.tr
0000-0002-4968-375x

Gökhan Akel
Antalya Belek University
gokhan.akel@belek.edu.tr
0000-0003-4353-7855

Bekir Değirmenci
Adıyaman Unıversıty
bdegirmenci@adiyaman.edu.tr
0000-0001-5236-5245

Dilek Aydoğdu
Karabuk Unıversıty
dilekaydogdu@karabuk.edu.tr
0000-0001-8042-7212

Kamile Elmasoğlu
Ankara Hacı Bayram Velı Unıversıty
kamile.elmasoglu@hbv.edu.tr
0000-0003-3811-3038

Hande Begüm Bumin Doyduk
Kocaelı Unıversıty
begumbumin@gmail.com
0000-0002-2917-2020

Arzu Şeker
Batman Unıversıty
arzu.seker@batman.edu.tr
0000-0002-3179-5956

Hatice Bahçe
Kayserı Unıversıty
haticebahce@kayseri.edu.tr
0000-0002-4358-4400

**Öz**

Merkezi bir otoriteye ihtiyaç duymayan ve blokzincir altyapısı ile güvenli dijital varlık transferlerine olanak tanıyan kripto paralar, giderek artan bir ilgiyle satın alınmaktadır. Küresel ve Türk yatırımcıların sayısındaki artışla, finansal dalgalanmalar karşısında bile dijital varlıklara olan ilginin sürdürülebilir bir şekilde artmaya devam edeceği açıktır. Ancak, tüketicilerin kripto para satın alırken blockchain teknolojisinin kullanım kolaylığı ve faydasını nasıl algıladığı henüz belirsizdir. Bu çalışma, kaliteli müşteri hizmeti, düşük maliyetler, verimlilik ve güvenilirlik gibi faktörleri göz önünde bulundurarak kripto para alımlarında blockchain teknolojisinin algılanan kullanım kolaylığı ve faydasını açıklamayı amaçlamaktadır. Bu hedefe ulaşmak için, Türkiye'nin farklı bölgelerinde kripto paralarla ilgilenen 463 katılımcıdan veri toplanmıştır. Veriler, SPSS Process Macro programları kullanılarak analiz edilmiştir. Analiz sonuçları, algılanan kullanım kolaylığı ve faydanın, müşteri hizmeti ve düşük maliyetler, verimlilik ve güvenlik üzerindeki etkilerini satın alma niyeti üzerinde aracılık ettiğini göstermektedir.

**Anahtar Kelimeler:** Blok Zinciri Teknolojisi, Kriptopara, Satın Alma Niyeti

## The Mediating Effect of Blockchain Technology on the Cryptocurrency Purchase Intention

**Abstract**

Cryptocurrencies, enabling secure digital asset transfers without a central authority, are experiencing increasing interest. With the increasing number of global and Turkish investors, it is evident that interest in digital assets will continue to rise sustainably, even in the face of financial fluctuations. However, it remains uncertain whether consumers perceive blockchain technology's ease of use and usefulness when purchasing cryptocurrencies. This study aims to explain blockchain technology's perceived ease of use and usefulness in cryptocurrency purchases by considering factors such as quality customer service, reduced costs, efficiency, and reliability. To achieve this goal, data were obtained from 463 participants interested in cryptocurrencies in different regions of Turkey. The data were analyzed using SPSS Process Macro programs. The analysis results indicate that perceived ease of use and usefulness mediate the effects of customer service and reduced costs, efficiency, and security on purchase intention.

## Introduction

Some traditional methods in various fields such as finance, logistics, marketing and production are giving way to digital innovations. With these digital innovations, cryptocurrencies and blockchain technology have recently started to attract attention. Digitalization has led to the emergence of digital currencies, gradually replacing traditional money in the economic system. Blockchain essentially represents a decentralized ledger system capable of documenting transactions involving two or more entities (Iansiti & Lakhani, 2017). The setup comprises a growing catalog of transactions that bear cryptographic signatures and cannot be undone. These transaction records are accessible to all individuals within a network. Consequently, every completed record includes a time mark and pointers to preceding transactions (Garg et al., 2021). This technology is a decentralized system that securely records and verifies transactions in a distributed digital ledger (Reid & Harrigan, 2013: 198). Blockchain technology-based cryptocurrencies facilitate the accurate recording and verification of transactions, facilitating the secure and fast transfer of digital assets through the blockchain infrastructure. The success of cryptocurrencies is largely based on the design principles of blockchain technology (Hashemi Joo et al., 2020).

Blockchain technology is considered promising and innovative and has a significant impact on consumers' cryptocurrency exchanges. Despite its volatility, the number of cryptocurrency owners has globally increased by 16.6% according to the Digital 2023 Global Overview Report. Turkey, in particular, has experienced a significant increase of 45.69% and ranks first in the country (Kemp, 2023). However, blockchain and cryptocurrency are still new technologies and future adoption within the financial system is difficult to predict.

Hence, this investigation holds significant value when viewed from a consumer standpoint, as it delves into its contribution to the act of acquiring cryptocurrencies through the utilization of blockchain technology. Within this context, the utilization of blockchain technology in cryptocurrency transactions prompts us to pose the subsequent research inquiries from a consumer vantage point.

RQ1. Is there an effect of perceived ease of use and usefulness for blockchain-based cryptocurrency purchase intention?

RQ2. What are the factors that affect the perceived usefulness and ease of use of the blockchain technology in the background of the crypto money transactions carried out by consumers?

The assessment of the technological adoption of blockchain technology has been explored in existing literature through various aspects like societal impact, structure, governmental endorsement, familiarity, reliance, uncertainty, ease, recognition of the brand, and caliber of the product. It is evident that the inclination to embrace cryptocurrencies based on blockchain is shaped by the perceived utility of these transactions (Albayati et al., 2020; Guych et al., 2018; Treiblmaier & Garaus, 2023).

At this point, the factors of whether consumers perceive quality customer service, reduced costs, efficiency, and reliability through blockchain technology when purchasing cryptocurrencies, and whether the ease of use and perceived usefulness mediate these effects form the starting point of the research. We use the widely used Technology Acceptance Model (TAM) to develop a model that can better predict the intention to use an innovative technology such as blockchain in cryptocurrency transactions. In numerous manners, our research adds to the existing body of knowledge. Primarily, comprehending the intermediary function of blockchain technology within the sphere of cryptocurrency acquisitions will unveil the extent of confidence investors place in blockchain technology and its influence on their inclinations when it comes to purchasing cryptocurrencies. Another contribution is to provide new findings on how the adoption of blockchain technology in the cryptocurrency market affects liquidity and investor interest. Additionally, it will provide critical information about investors' perceptions and behaviors regarding the transparency and security features provided by blockchain technology, shaping the future of the sector and regulatory frameworks. The results of this research will offer valuable perspectives for the advancement of forthcoming cryptocurrency systems, enlightening researchers, investors, and professionals engaged in domains such as commerce, investment, economics, and business frameworks. Furthermore, the methodology and analysis methods of the research are thought to be useful to fill the gaps in the literature and open up new fields of study by inspiring further studies on blockchain technology and crypto money research.

This study examines the factors influencing the perceived usefulness and ease of use of blockchain technology among consumers who purchase cryptocurrencies, using the TAM. The ongoing development and innovation in blockchain-based cryptocurrencies are expected to attract researchers, investors, and experts, fostering advancements in areas such as trade, investment strategies, economic models, and business practices.

Our research is organized as follows. First of all, the concepts of cryptocurrency, blockchain and studies in the literature are explained. subsquently, utilizing the TAM model as the conceptual framework, the basis of theory from prior studies is employed to assess the perceived simplicity and advantages of applying blockchain technology. Following this, hypotheses are formulated and substantiated. Subsequently, particulars concerning data compilation and analysis are provided in the methodology segment. In the concluding portion, the discussion chapter elaborates on our discoveries and outlines the theoretical as well as pragmatic inputs of the study. Ultimately, the study's limitations and prospects for future endeavors are articulated.

## Literature Review

### Cryptocurrency

Cryptocurrencies and their transformative powers have been widely discussed and received considerable attention. The increasing demand for cryptocurrencies can be attributed to price fluctuations that have generated significant interest among investors and users. Cryptocurrencies and virtual payment methods have emerged as innovative alternatives to traditional financial systems, providing new avenues for conducting transactions and facilitating economic interaction in today's digital economy. The utilization of blockchain technologies has enabled the implementation of virtual methods and the rise of cryptocurrencies as a significant phenomenon (Garcia-Monleon et al., 2021).

Today, with the rapid rise in digitalization, cryptocurrencies are used as investment tools, and even significant investments can be made in these currencies on a global scale (Yılmaz & Ecemiş, 2022). Bitcoin, introduced to the world through a paper published by Satoshi Nakamoto in 2008, has played a pioneering role in the development of cryptocurrencies. Bitcoin, free from centralized control, is highly volatile (Gupta & Bagga, 2017). The first scientific study on cryptocurrencies focused on Bitcoin (Jacob, 2011). Subsequent research focused on analyzing the price dynamics and transaction patterns of Bitcoin (Gronwald, 2014; Cheah & Fry, 2015; Gupta, 2017) and the relationship between cryptocurrency pricing and assets such as gold, silver, and oil (Zimmer, 2017; Sukamulja & Sikora, 2018). In the context of price-oriented studies, speculative assets in the cryptocurrency market exhibit high volatility and behave independently of external economic factors, based on the principles of supply and demand (Garcia-Monleon et al., 2021).

Determining the economic impact of cryptocurrencies and understanding their internal pricing mechanisms under current market conditions are essential aspects of research. Many studies have primarily explored the conceptual aspects of cryptocurrencies, such as their technological foundations, regulatory challenges, and potential implications for financial systems. However, cryptocurrencies can be viewed as consumption tools and virtual currencies that possess objectivity or value as commodities (Abboushi, 2017). From another perspective, cryptocurrencies represent the value of blockchain technology and have inherent valuations in the market (Engelhardt, 2017).

The cryptocurrency market is heterogeneous and decentralized. While Bitcoin initially emerged as the primary and standard currency within this market, alternative cryptocurrencies and their users/providers have emerged, shaping the cryptocurrency market. This study specifically investigates three key dimensions of cryptocurrency usage, focusing on economic purposes, mobile trading, and cross-network asset transfers, to provide empirical insights. The first dimension focuses on specific economic purposes, while the second involves mobile trading and the transfer of cryptocurrencies within the blockchain world. The third dimension is multi-layered, enabling asset transfers between networks by employing cryptocurrencies for multiple purposes and functionalities in the market (Garcia-Monleon et al., 2021).

**Blockchain Technology**

Cryptocurrencies and blockchain technologies are closely intertwined. Blockchain technology serves as the underlying infrastructure for cryptocurrencies and enables the consolidation of disparate and disparate elements to address cybersecurity issues. It supports asset protection over the network by leveraging Internet of Things (IoT) tools. Blockchain technology eliminates threats and potential disruptions in a system, making it a smart and secure system. Transactions within the system were conducted online (Dorri et al. 2017).

Cryptocurrencies, particularly Bitcoin, are often considered when discussing blockchain technology. A blockchain is viewed as an outcome of Bitcoin and is defined as a chain of blocks or a system that connects to a previous block. Although there was substantial curiosity in this novel technological domain at the outset, demand has experienced a decline since 2018 due to the prudent uptake of blockchain technology by businesses (Aggarwal et al., 2022). This data stack is processed in the technological field, forming interconnected information blocks to create a "chain" within the system (Minoli & Occhiogrosso, 2018). A blockchain system securely stores data records in a decentralized manner and acts as a digital ledger (Khan & Salah, 2018).

Blockchain technology does not rely on physical payment methods or money. Instead, it encompasses the various cryptocurrencies used in digital transactions. Cryptocurrencies facilitate secure digital payment methods and offer opportunities for online shopping within a decentralized network (Raymaekers, 2015). Users perceive cryptocurrencies as financial investment tools. Bitcoin is the most well-known and commonly purchased cryptocurrency globally, making it valuable. According to a study conducted by the World Economic Forum (WEF) in 2021, developing countries rank at the top in terms of cryptocurrency adoption rates worldwide. Nigeria holds the first position at 33%, followed by Vietnam at 21% in second place, the Philippines at 20% in third place, and Turkey at 16% in fourth place. However, Japan currently has relatively low cryptocurrency usage, but there is a growing preference for domestic cryptocurrencies in the near future (WEF, 2021).

Recent studies have focused on information technologies and their utilization, emphasizing the importance of assessing individuals' and groups' attitudes towards knowledge-based systems (Lai & Li, 2005). TAM has made significant theoretical and experimental contributions to understanding user acceptance of information technologies and measuring their attitudes (Davis et al., 1989). TAM allows for the analysis of individuals' reasons for accepting technology, their resistance, views on innovation, and theoretical and practical experiences (Adams et al., 1992). While the TAM was initially developed to predict the behavior of potential users with no prior experience in information technologies, variations in behavior and perception may arise between knowledgeable and potential users (Özer et al., 2010). The framework unveiled the impacts of recognized advantages on mindsets and intentions for action, with perceived utility being the central factor, trailed by perceived simplicity. The connection between ease of use and perceived utility still requires further elucidation. Attitude effectively shapes the intention for action and the approach through which a conduct is executed (Lee et al., 2007). TAM has been effectively adjusted to scrutinize user reception of online commercial activities, the incorporation of the Internet in learning, and diverse categories of information technology (Torres et al., 2006; Shen & Eder, 2009).

**Theoretical Background**

Approaches to modeling devised for comprehending the acceptance and uptake of technology have found application across various fields (Carter & Belanger, 2005: 8). TAM is recognized as the most proficient framework for elucidating users' adoption of technologies due to its extensive array of applications and widespread utilization (Lee et al., 2003). Originated by Davis (1989), TAM is founded upon a revision of the Theory of Reasoned Action (Fishbein & Ajzen, 1975) and stands as the prevailing model for evaluation. The premise of this model posits that when users confront novel technology, their decisions on the how and when of its usage are influenced by a set of factors. The theory hinges on two main tenets: perceived utility and simplicity of use (Lindsay et al., 2011).

The research framework and suppositions, rooted in theoretical underpinnings, are constructed on a structural schema using information derived from scholarly sources. The choice of TAM as the theoretical basis for the research model stemmed from its alignment with the research objectives and queries. The model was expanded by integrating external variables like exceptional customer service, cost reduction, efficiency, and security aspects. In this scenario, the core elements of TAM—perceived utility, perceived ease of use, attitude, and intention—are sequentially expounded.

Perceived utility signifies an individual's conviction that employing a specific system or technology will augment job performance, while perceived ease of use denotes an individual's belief that using a system is uncomplicated and necessitates minimal exertion (Davis, 1989: 320). Attitude embodies an individual's sentiments and sentiments toward a situation, encompassing favorable or adverse feelings concerning the intended behavior. Intention gauges an individual's readiness to partake in the envisaged behavior (Fishbein & Ajzen, 1975: 216). Diverse investigations have accentuated that TAM can be extended to grasp the acceptance of assorted technologies. Consequently, it is posited that TAM can be proficiently employed to scrutinize intentions toward the use of blockchain technology.

**Hypotheses Development**

Crafted by Davis et al. (1986: 985-986), the TAM is designed to delve into individuals' inclinations to embrace technology. The framework elucidates that external factors shape perceived simplicity and value. Perceived value shapes users' attitudes and intentions regarding technology use, while perceived simplicity affects both perceived value and users' attitudes regarding technology uptake. Earlier investigations suggest that the caliber of customer service yields a favorable influence on perceived value. Zeithaml and Bitner (2003) demonstrated that quality customer service enhances users' perception of a product or service's usefulness, and contributes to customer satisfaction. Similarly, Grönroos (2007) found that the quality of customer service increases users' perception of a product or service's usefulness, and positively influences customer satisfaction. Based on these findings, the following hypothesis was developed:

**H$_{1a}$.** Quality customer service positively and significantly influences perceived usefulness in blockchain technology.

Meuter et al. (2000) found that quality customer service reduces the difficulties users may encounter during product or service usage, thereby increasing the perceived ease of use. Bei & Chiao (2001) showed that service quality is an important factor in measuring consumer satisfaction and loyalty through perceived ease of use. Cho & Sagynov (2015), in their study applying the Technology Acceptance Model, found that service quality positively affects perceived ease of use. Hanjaya et al. (2019) examined the impact of service quality on consumers' behavior in Indonesia and Singapore. The study found that service quality significantly affects online purchasing intention via mobile application through perceived ease of use. Therefore, the following hypothesis was developed:

**H$_{1b}$.** Quality customer service positively and significantly influences the perceived ease of use of blockchain technology.

In the context of a blockchain-based platform, quality customer service can enhance user experience and trust, ultimately influencing the purchase intention. Quality customer service can provide users with prompt and effective support, resolve issues, improve security measures, and provide a secure environment. As a result, users' trust in the blockchain platform and their positive user experience can lead to an increase in their purchase intention (Liu et al. 2023). Therefore, the following hypothesis is proposed:

**H$_{1c}$.** Quality customer service using blockchain technology has a positive and significant effect on purchase intention.

Diminished expenses pertain to the reduction or eradication of costs linked with the utilization of a product or service. Research has shown that reducing costs can enhance user perceptions of a product or service's usefulness. Kim and Kankanhalli (2009) noted that reducing cognitive load can increase users' perception of usefulness. Reduced costs can reduce the costs that users may encounter when performing tasks, thereby reducing the cognitive load and leading users to perceive the product or service as more useful. Similarly, Liang and Lai (2002) find that reducing the costs associated with using a product or service can increase users' perceived usefulness. Therefore, we propose the following hypothesis:

**H$_{2a}$.** Reduced costs positively and significantly influence the perceived usefulness of the blockchain technology.

Reduced costs have always been a significant factor in cost reduction and technology adoption. Blockchain technology can reduce costs (Garg et al. 2022: 4). Reduced costs involve reduced time, emotional effort, and money (Ho & Ko, 2008). Lai and Lin (2016) found that reduced costs help users perceive a product or service as easier to use. Reduced costs can lower the expenses that users may encounter when using a product or service. Grewal et al. (2017) emphasized that reduced costs can alleviate the difficulties users face during product or service usage, thus enhancing perceived ease of use. Reducing the costs associated with using a product or service can help users perceive the product or service as easier to use. Therefore, we propose the following hypothesis:

**H$_{2b}$.** Reduced costs positively and significantly influence the perceived ease of use of blockchain technology.

Coco et al. (2017: 6) stated that blockchain technology can be managed much more efficiently than traditional systems in terms of transaction speed and cost reduction. Reduced cost can positively influence customers' purchase intentions. Lower prices, discounts, promotions, and other cost-reducing factors offer customers tangible advantages. Customers evaluate the benefits of shopping in an environment in which more cost-effective products or services are offered, which may increase their purchase intention. Reduced costs also create opportunities to provide customers with more value. Lower-cost products or services may be more affordable for customers' budgets and more effective in meeting their demands, which can increase their purchase intention (Esfahbodi et al. 2022; Pandey et al. 2019). Therefore, the following hypothesis is proposed:

**H$_{2c}$.** The reduced cost of blockchain technology has a positive and significant effect on the purchase intention.

Customer trust, which reduces risk and uncertainty, is a crucial factor in individuals' decisions to develop long-term relationships with service providers, brands, and products (Gao & Bai, 2014). Both efficiency and security factors contribute to the perceived usefulness of a product or service. Davis et al. (1989) found that efficiency is one of the key elements of usefulness and that a product or service's efficiency enhances users' perception of usefulness. Security factors can also influence user perceptions of their usefulness. Research has shown that users trust products or services without more security vulnerabilities and that this trust increases their perception of usefulness. Therefore, the following hypotheses are developed and tested:

**H$_{3a}$.** Efficiency and security positively and significantly influence the perceived usefulness of the blockchain technology.

Efficiency is defined as the ability of a product or service to quickly and effectively perform a specific task. Security, on the other hand, relates to users' sense of safety and minimization of risks. Liu et al. (2019) found that an efficient interface enhances users' perception of ease of use, leading to increased user satisfaction. Alalwan et al. (2018) found that users' perception of security positively influences their evaluation of website usability. Based on these studies, the following hypothesis was developed:

**H$_{3b}$.** Efficiency and security positively and significantly influence the perceived ease of use of blockchain technology.

Mak and Ramayah (2017) and Yousafzai et al. (2010) have examined how perceived benefits, ease of use, enjoyment factors, and trust influence purchase intention in various contexts, including online shopping and mobile banking. Trust in disruptive technologies is a crucial factor for customer loyalty (Ullah et al. 2022:6). Folkinshteyn and Lennon (2016) note that despite the high risks associated with cryptocurrency service providers, the reliability and robustness of blockchain attract new users. Based on the results of these studies, the following hypothesis is proposed:

**H$_{3c}$.** Efficiency and security in blockchain technology have positive and significant effects on purchase intention.

Yeh et al. (2019) illustrate that blockchain technology possesses the capability to establish security and impact the intention to purchase. Mendoza-Tello et al. (2018) reveal that the perceived value of blockchain technology has repercussions on the intent to purchase. The utilization of social commerce and social support impacts both the perceived value of blockchain technology and the intention to purchase. Through elements such as security, genuineness, advantages, and user-friendliness, blockchain technology can elevate consumer attitudes and thereby enhance the intention to purchase. Given the profoundly inventive and technology-driven essence of the process of adopting cryptocurrencies, it is suitable to employ TAM variables, encompassing perceived simplicity and perceived value. Hence, we posit the subsequent hypothesis:

**H$_4$.** The perceived usefulness of blockchain technology has a positive and significant effect on purchase intention.

Albayati et al. (2020) found in their research that the perceived ease of use in blockchain technology influences purchase intention in cryptocurrencies. The study, which was conducted through the integration of the TAM and the Expectation-Confirmation Model (ECM), demonstrates that an increase in the perceived ease of use of blockchain technology positively affects purchase intention. Teo, Lim, and Lai (2019) revealed in their research that the perceived ease of use of blockchain technology influences purchase intention in cryptocurrencies. Intrinsic and extrinsic motivational factors also affect purchase intentions. Perceived ease of use enhances the intention to use cryptocurrencies. A experimental study conducted by Kim, Kim, and Kwon (2019) showed that the perceived ease of use of blockchain technology has a positive effect on purchase intention in cryptocurrencies. The experiment revealed that individuals who believed blockchain technology to be easy to use had an increased purchase intention. The common conclusion drawn from these articles is that the perceived ease of use of blockchain technology positively influences purchase intention in cryptocurrencies. As users perceive blockchain technology as easily usable, their purchase intentions increase. Perceived ease of use is a motivating factor for cryptocurrency usage, and supports users in adopting this technology.

**H5.** Perceived ease of use of blockchain technology has a positive and significant effect on purchase intention.

Davis (1989) stressed the significance of perceived value and perceived simplicity in the utilization of information technology. Perceived value denotes a user's assessment of the advantages offered by the technology, while perceived simplicity concerns how effortlessly and effectively the user can engage with the technology. This investigation illustrates that perceived value exerts a favorable influence on perceived simplicity. Venkatesh et al. (2019) observed that perceived value shapes users' acceptance and intention to engage with a technology. Perceived value embodies a user's conviction that technology can enable them to employ it proficiently and achieve distinct objectives. Consequently, perceived value can impact a user's perspective on the user-friendliness of a product or service. Chen et al. (2002) centered on the effect of perceived value on users' uptake of products or services in the online sphere. Perceived value encompasses the users' belief that the online encounter will streamline and be advantageous for them. This exploration disclosed that perceived value favorably affects perceived simplicity, subsequently amplifying user contentment. Based on these findings, we posit the ensuing hypotheses:

**H6.** Perceived usefulness in blockchain technology has a positive and significant effect on its perceived ease of use in blockchain technology.

Blockchain, considered an emerging disruptive technology, offers benefits ranging from meeting businesses' legal requirements to reducing transaction costs; enhancing efficiency and security; providing transparent, fast, and high-quality service delivery; and increasing customer satisfaction (Garg et al., 2020). Cryptocurrencies are the most commonly used mechanisms in the blockchain technology. Despite being relatively new, both blockchain technology and the adoption of cryptocurrencies raise concerns such as user privacy and trust. The first stage in adopting blockchain technology is the adoption of cryptocurrencies (McDougall, 2014; Sohaib et al., 2020: 13139-13140). In blockchain technology, quality customer service and reduced cost, efficiency, and security can help users better understand the technology, resolve issues, enhance the user experience, and consequently influence purchase intention. This can increase the users' perceived usefulness levels. In this context, we propose the following hypothesis:

**H7a.** The mediating effect of the perceived usefulness of blockchain technology exists in the positive and significant influence of quality customer service on purchase intention in cryptocurrencies.

**H7b.** The mediating effect of the perceived usefulness of blockchain technology exists in the positive and significant influence of reduced costs on purchase intention in cryptocurrencies.

**H7c.** The mediating effect of the perceived usefulness of blockchain technology has a positive and significant influence of efficiency and security on the purchase intention of cryptocurrencies.

The perceived of ease of use, devised to evaluate a user's anticipation of a system's effortlessness and absence of complexity, holds a pivotal function in determining the user's endorsement of the system. When users perceive a system as straightforward to operate, their likelihood of embracing it increases. Moreover, when the attributes of a system align with task requisites, the perceived value of the system is anticipated to amplify the user's job performance (Davis 1989; Gefen et al. 2003; Van der Heijden et al. 2003; Jen et al. 2009). Investigations indicate that ease of use impacts the perceived value of a technology and wields a substantial influence on users' intention to employ it (Davis 1989; Venkatesh et al. 2003; Lin and Chang 2011; Koivisto et al. 2016; Compernolle et al. 2018). Chen et al. (2018) scrutinize the effect of high-quality customer service on perceived simplicity and its association with the intention to purchase. Their exploration delved into how premium customer service affects the perceived simplicity of e-wallet users and, consequently, their intent to make a purchase. Grounded in the foundation laid by these investigations, the ensuing hypothesis was postulated:

**H7d.** The mediating effect of perceived ease of use in blockchain technology exists in the positive and significant influence of quality customer service on purchase intentions in cryptocurrencies.

**H$_{7e}$.** The mediating effect of perceived ease of use in blockchain technology exists in the positive and significant influence of reduced costs on purchase intention in cryptocurrencies.

**H$_{7f}$.** The mediating effect of perceived ease of use of blockchain technology has a positive and significant influence of efficiency and security on the purchase intention of cryptocurrencies.

Both in digital and physical realms, dealing with users' lack of confidence and fostering reliance are pivotal. Each individual necessitates a distinct level of assurance to partake in online activities and advance. Despite the Internet's recognition as an efficient and swift tool for both enterprises and consumers, the significance of exceptional customer service and contentment has undergone a notable surge. Dehghanpuri et al. (2020) scrutinize the connections between internal variables like customer contentment and external variables such as service quality, privacy, and trust, employing a structured equation modeling methodology. Their inquiry divulged that furnishing top-notch customer service culminates in customer satisfaction and is influenced by trust in the domain of online services. Ullah et al. (2022: 9) ascertain that cost reductions yield a favorable impact on perceived simplicity, perceived value, and the intention to utilize. Moreover, they identified that while trust exerts a significant sway on perceived simplicity and usage intent, its influence on perceived value is negligible.

Users perceive a system as easy when they do not require extensive training to learn or use it without effort, and when they consider the system to be highly beneficial for their work (Dasgupta et al., 2002:89). This perception of ease of use leads users to accept the system as user friendly. Consequently, users expect to use the system to enhance effectiveness, job performance, work quality, and productivity (Staples et al., 2002:118) while also achieving cost savings, increasing service quality without increasing costs, enhancing production without increasing costs, and coping with change (Whyte et al., 1997:38; Hartman et al., 2002:929; Legris et al., 2003:191). Several studies have examined the relationships among perceived ease of use, perceived usefulness, and trust, suggesting that perceived ease of use and perceived usefulness have a positive impact on trust and, consequently, a significant effect on usage intention (Gefen et al., 2003; Nicolaou and McKnight, 2006; Horst et al., 2007; Belanche et al., 2012; Schnall et al., 2015).

Chiou and Droge (2006: 624) emphasized the importance of customer satisfaction and trust in increasing behavioral loyalty by directly or indirectly convincing consumers to invest in specific assets. Ko et al. (2018) and Ullah et al. (2020) demonstrated that cost reduction positively influences perceived ease of use. Chaveesuk et al. (2020: 140) state that trust, preparedness, and resources play a crucial role in influencing acceptance and behavioral intention by encouraging perceived financial cost and facilitating conditions.

This study aims to evaluate the mediating effects of perceived ease of use and perceived usefulness, which are independent factors from the TAM, on quality customer service, reduced cost, security, and efficiency, which are external variables that influence purchase intention in cryptocurrencies within the context of blockchain technology. Based on this background, the following hypothesis is proposed.

**H$_{8a}$.** The sequential mediating effect of perceived ease of use and perceived usefulness of blockchain technology exists in the positive and significant influence of quality customer service on purchase intention in cryptocurrencies.

**H$_{8b}$.** The sequential mediating effect of perceived ease of use and perceived usefulness in blockchain technology exists in the positive and significant influence of reduced costs on purchase intention in cryptocurrencies.

**H$_{8c}$.** The sequential mediating effect of perceived ease of use and perceived usefulness of blockchain technology exists in the positive and significant influence of efficiency and security on purchase intention in cryptocurrencies.

### Methodology

This study aims to investigate the perceived usefulness and perceived ease of use of blockchain technology in consumers' cryptocurrency purchases from the perspective of the TAM. In this regard, perceived usefulness and perceived ease of use are thought to mediate the potential impact of quality customer service, low cost, efficiency, and reliability on cryptocurrency purchase intentions. Based on a literature review, a research model depicting the relationships between the variables was developed. The proposed model, along with 21 hypotheses, is grounded in the literature. To test the hypotheses, analysis was conducted using Process Macro 4.2 program developed by Hayes (2018), and integrated as an add-on in the SPSS software. Specifically, Process Macro 4.2 version was utilized for simple, multiple, and sequential regression analyses. The study included six variables: three independent variables (quality customer service, reduced cost, efficiency, and reliability), two mediator variables (perceived usefulness and perceived ease of use), and one dependent variable (purchase intention). The research model is illustrated in Figure 1.

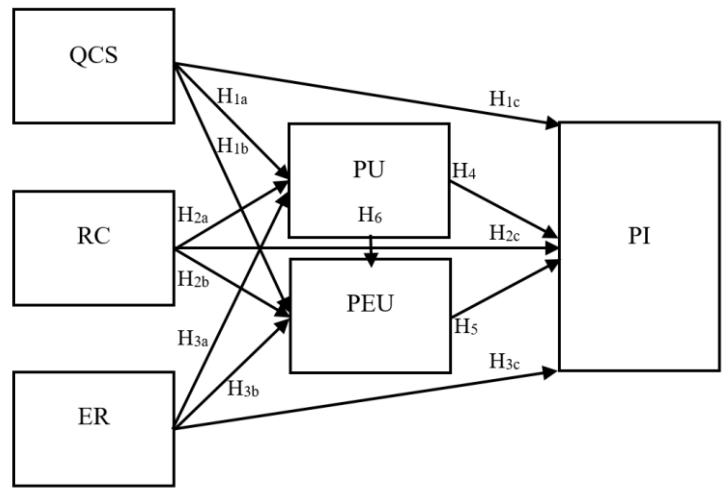

**QCS:** Quality Customer Service; **RC:** Reduced Costs; **ES:** Efficiency and Security;

**PU:** Perceived Usefulness; **PEU:** Perceived Ease of Use; **PI:** Purchase Intention

**Figure 1.** Research Model

## Measurement Development

A survey was developed by adapting measurement scales from previous studies, specifically those developed by Garg et al. (2021), Lopez & Shih (2023), and Ullah et al. (2022). Because the survey was intended for use in Turkey, a proficient translator translated the original English version into Turkish. Subsequently, the Turkish questionnaire was back translated into English to ensure translation accuracy and consistency.

## Data Collection and Sample

Data were collected using Google Forms from 482 participants; however, 19 participants were excluded from the analysis because of inaccurate or incomplete data, resulting in a final sample size of 463. Participants were initially screened to determine their interest in purchasing cryptocurrencies, and if affirmative, they proceeded to answer additional questions. The data collection process consisted of two parts. The first part included demographic questions regarding the participants' gender, marital status, age, occupation, education, and income level. The second part consisted of questions related to the quality of customer service, reduced cost, efficiency and reliability, perceived usefulness, perceived ease of use, and purchase intention. These questions were divided into six sub-dimensions comprising a total of 29 items. Convenience sampling was employed to gather data from individuals aged 18 and above residing in different regions of Turkey who had an interest in cryptocurrency. Participants were initially screened to determine their interest in purchasing cryptocurrencies, and if affirmative, they proceeded to answer additional questions. The data collection process consisted of two parts. The first part included demographic questions regarding the participants' gender, marital status, age, occupation, education, and income level. The second part consisted of questions related to the quality of customer service, reduced cost, efficiency and reliability, perceived usefulness, perceived ease of use, and purchase intention. These questions were divided into six sub-dimensions; quality customer service, efficiency and reliability, perceived usefulness, perceived ease of use, and purchase intention, and four items each for reduced cost. All questions in this section were rated on a five-point Likert scale ranging from 1 (strongly disagree) to 5 (strongly agree). The data collection process involved a pilot study. Prior to the main study, a pilot test was conducted with 48 participants to assess the questionnaire's suitability. The pilot test evaluated the clarity, reliability, and validity of the questionnaire, and the results indicated that the questions were well understood, and that the questionnaire was suitable for analysis.

Table 1. Survey Items

| Variable | Items | Mean | Sd |
|---|---|---|---|
| QCS | Blockchain technology enhances transparency in cryptocurrency transactions | 3.72 | 1.327 |
| | Blockchain technology instills trust in cryptocurrency transactions | 3.74 | 1.308 |
| | Blockchain technology improves data accuracy in cryptocurrency transactions | 3.72 | 1.346 |
| | Blockchain technology mitigates risks in cryptocurrency transactions | 3.81 | 1.332 |
| | Blockchain technology automates actions and transactions between parties in cryptocurrency transactions | 3.61 | 1.355 |
| RC | Blockchain technology reduces transaction costs in cryptocurrency transactions | 3.65 | 1.297 |
| | Blockchain technology eliminates intermediaries in cryptocurrency transactions | 3.62 | 1.247 |
| | Blockchain technology reduces administrative costs in cryptocurrency transactions | 3.60 | 1.256 |
| | Blockchain technology reduces operational costs in cryptocurrency transactions | 3.64 | 1.252 |
| ER | Blockchain technology helps monitor real-time cryptocurrency transactions | 3.22 | 1.246 |
| | Blockchain technology increases the speed of cryptocurrency transactions | 3.39 | 1.259 |
| | Blockchain technology enhances efficiency in cryptocurrency transactions | 3.30 | 1.258 |
| | Blockchain technology enhances security in cryptocurrency transactions | 3.38 | 1.266 |
| | Blockchain technology improves system integrity in cryptocurrency transactions | 3.34 | 1.265 |
| PU | Using the blockchain infrastructure makes it easy to use cryptocurrencies | 4.02 | 1.396 |
| | Learning how to use cryptocurrencies with the blockchain infrastructure is easy | 3.99 | 1.389 |
| | Understanding the interaction of cryptocurrencies with the blockchain infrastructure is easy | 4.02 | 1.432 |
| | Finding information about cryptocurrencies using blockchain infrastructure is easy | 4.11 | 1.374 |
| | The process of buying cryptocurrencies is easy with the blockchain infrastructure | 4.04 | 1.361 |
| PEU | Trading cryptocurrencies is easy with the blockchain infrastructure | 4.17 | 1.356 |
| | Trading cryptocurrencies with blockchain infrastructure is fast | 4.15 | 1.284 |
| | The blockchain infrastructure can improve my financial and business performance through cryptocurrencies | 4.12 | 1.348 |
| | The blockchain infrastructure can enhance my financial and commercial performance through cryptocurrencies | 4.12 | 1.308 |
| | The features of crypto money and the blockchain infrastructure are useful for me | 4.14 | 1.357 |
| PI | I am planning to purchase cryptocurrencies | 3.55 | 1.352 |
| | I would recommend cryptocurrencies to my friends, family, and acquaintances | 3.49 | 1.366 |
| | I am not hesitant to provide information about cryptocurrencies | 3.47 | 1.373 |
| | I use my credit card to purchase cryptocurrencies | 3.44 | 1.384 |
| | In the future, I will conduct my transactions using cryptocurrencies | 3.72 | 1.353 |

## Results

Univariate normality was assessed using skewness-kurtosis tests as outlined by Hair et al. (2010) and Kline (2011). This study scrutinized skewness and kurtosis values to inspect the distribution's normality, as suggested by Tabachnick and Fidell (2013). The normality of the distribution was tested using skewness and kurtosis values. The analysis revealed that the kurtosis value ranged from -1.369 to -0.997, while the skewness value fell between 0.255 and 1.474. Given that both skewness and kurtosis values fell within the -2 and +2 ranges, it was concluded that the data followed a normal distribution (George and Mallery, 2010). Additionally, both the Multiple Normality Distribution and Kolmogorov-Smirnov tests confirmed the data's adherence to a normal distribution.

### Descriptive Statistics

The study included 463 participants, with slightly more men (249) than women (214). The majority of participants (57.5%) were single, while 42.5% were married. The age distribution was relatively even among participants under the age of 45 years, with similar numbers in the 18-24, 25-34, and 35-44 age groups. In terms of education, most participants held qualifications higher than an associate's degree, including bachelor's and post-graduate degrees. The majority of participants were employed in either the public or private sector. The majority reported monthly incomes ranging from 18001 TL-27000 TL. Further details are provided in Table 2.

Table 2. Frequency Analysis by Demographic Variables

| Demographic Variable | Categories | N | % |
|---|---|---|---|
| Gender | Female | 214 | 46.2 |
| | Male | 249 | 53.8 |
| Marital Status | Single | 266 | 57.5 |
| | Married | 197 | 42.5 |
| Age | 18-24 | 115 | 24.8 |
| | 25-34 | 133 | 28.7 |
| | 35-44 | 133 | 28.7 |
| | 45-54 | 44 | 9.5 |
| | 55-64 | 32 | 6.9 |
| | 65 and over | 6 | 1.3 |
| Occupation and Employment Status | Student | 85 | 18.4 |
| | Public Sector | 106 | 22.9 |
| | Private Sector | 183 | 39.5 |
| | Business Owner | 32 | 6.9 |
| | Housewife | 16 | 3.5 |
| | Retired | 21 | 4.5 |
| | Unemployed | 20 | 4.3 |
| Education Status | Primary School | 30 | 6.5 |
| | High School | 77 | 16.6 |
| | Associate Degree | 124 | 26.8 |
| | Bachelor's Degree | 187 | 40.4 |
| | Graduate Degree | 45 | 9.7 |
| Income | 9000 TL and below | 132 | 28.5 |
| | 9001 TL-18000 TL | 53 | 11.4 |
| | 18001 TL-27000 TL | 168 | 36.3 |
| | 27001 TL-36000 TL | 71 | 15.3 |
| | 36001 TL and above | 39 | 8.4 |

**Measurement Model**

Exploratory factor analysis was initially applied to the collected data. The Kaiser-Meyer-Olkin (KMO) measure of sampling adequacy (0.955) and Bartlett's test of sphericity ($p < 0.001$) indicated that factor analysis was appropriate. According to the results of the exploratory factor analysis presented in Table 3, the items loaded onto the six factors, as expected. The total variance explained by these factors was calculated to be 80.39%. The questions were divided into six sub-dimensions comprising a total of 29 items: five items each for quality customer service, efficiency and reliability, perceived usefulness, perceived ease of use, and purchase intention, and four items each for reduced cost.

Table 3. Exploratory Factor Analysis

| Variable | 1 | 2 | 3 | 4 | 5 | 6 |
|---|---|---|---|---|---|---|
| QCS1 | .771 | | | | | |
| QCS2 | .803 | | | | | |
| QCS3 | .773 | | | | | |
| QCS4 | .780 | | | | | |
| QCS5 | .717 | | | | | |
| RC1 | | .798 | | | | |
| RC2 | | .827 | | | | |
| RC3 | | .845 | | | | |
| RC4 | | .705 | | | | |
| ER1 | | | .742 | | | |
| ER2 | | | .725 | | | |
| ER3 | | | .743 | | | |
| ER4 | | | .700 | | | |
| ER5 | | | .752 | | | |
| PU1 | | | | .868 | | |
| PU2 | | | | .860 | | |
| PU3 | | | | .866 | | |
| PU4 | | | | .858 | | |
| PU5 | | | | .865 | | |
| PEU1 | | | | | .840 | |
| PEU2 | | | | | .849 | |
| PEU3 | | | | | .841 | |
| PEU4 | | | | | .846 | |
| PEU5 | | | | | .836 | |
| PI1 | | | | | | .797 |
| PI2 | | | | | | .827 |
| PI3 | | | | | | .856 |
| PI4 | | | | | | .814 |
| PI5 | | | | | | .785 |

## Reliability and Validity

The scales' validity and reliability were examined through internal consistency (Cronbach's alpha coefficient), convergent validity, and discriminant validity. Reliability and validity analyses of the scale were conducted using Cronbach's alpha (CA), construct reliability (CR), and average variance extracted (AVE). The overall Cronbach's alpha for the 29 items was approximately 0.926, indicating good internal consistency and reliability. The results are presented in Table 4. The reliability of the scales was assessed using Cronbach's alpha coefficient and Composite Reliability (CR) values, following the methodology outlined by Hair et al. (2017). The Cronbach's alpha values for the scales range from 0,880 to 0,917. According to Hair et al. (2017), alpha values exceeding 0,70 are sufficient for internal consistency. Average Variance Extracted (AVE) values are calculated for convergent validity, and values exceeding 0,50 are considered acceptable (Hair et al., 2017). All variables except for the belief dimension of religiosity meet this criterion. However, convergent validity is acceptable due to the belief dimension's AVE value being close to 0,50 and meeting other measures such as Factor Loadings, CR, and Alpha. In summary, the reliability and convergent validity of the scales are generally in line with the recommended standards set by Hair et al. (2017).

Table 4. CA, AVE, and CR Values

| Constructs | CA | AVE | CR |
|---|---|---|---|
| QCS | .901 | .592 | .879 |
| RC | .917 | .679 | .894 |
| ER | .888 | .537 | .853 |
| PU | .880 | .746 | .936 |
| PEU | .970 | .710 | .924 |
| PI | .913 | .666 | .909 |

Model 6, developed by Hayes (2018), was used for mediation, moderation, and conditional process analyses. Details of the hypotheses, paths, and regression analyses are presented in Table 5.

Table 5. Information regarding Hypotheses and Regression Analysis

| Hypothesis No | Relationships | Simple Regression | Multiple Regression | Sequential Regression |
|---|---|---|---|---|
| $H_{1a}$ | QCS $\Longrightarrow$ PU | X | | |
| $H_{2a}$ | RC $\Longrightarrow$ PU | X | | |
| $H_{3a}$ | ER $\Longrightarrow$ PU | X | | |
| $H_{1b}$ | QCS $\Longrightarrow$ PEU | X | | |
| $H_{2b}$ | RC $\Longrightarrow$ PEU | X | | |
| $H_{3b}$ | ER $\Longrightarrow$ PEU | X | | |
| $H_6$ | PU $\Longrightarrow$ PEU | X | | |
| $H_{1c}$ | QCS $\Longrightarrow$ PI | X | | |
| $H_{2c}$ | RC $\Longrightarrow$ PI | X | | |
| $H_{3c}$ | ER $\Longrightarrow$ PI | X | | |
| $H_4$ | PU $\Longrightarrow$ PI | X | | |
| $H_5$ | PEU $\Longrightarrow$ PI | X | | |
| $H_{7a}$ | QCS $\Longrightarrow$ PU $\Longrightarrow$ PI | | X | |
| $H_{7b}$ | RC $\Longrightarrow$ PU $\Longrightarrow$ PI | | X | |
| $H_{7c}$ | ER $\Longrightarrow$ PU $\Longrightarrow$ PI | | X | |
| $H_{7d}$ | QCS $\Longrightarrow$ PEU $\Longrightarrow$ PI | | X | |
| $H_{7e}$ | RC $\Longrightarrow$ PEU $\Longrightarrow$ PI | | X | |
| $H_{7f}$ | ER $\Longrightarrow$ PEU $\Longrightarrow$ PI | | X | |
| $H_{8a}$ | QCS $\Longrightarrow$ PU $\Longrightarrow$ PEU $\Longrightarrow$ PI | | | X |
| $H_{8b}$ | RC $\Longrightarrow$ PU $\Longrightarrow$ PEU $\Longrightarrow$ PI | | | X |
| $H_{8c}$ | ER $\Longrightarrow$ PU $\Longrightarrow$ PEU $\Longrightarrow$ PI | | | X |

Regression analyses were conducted in three stages, following the procedures outlined in Model 6. In the first stage, simple, multiple, and sequential regression analyses were performed for independent variable 1 (X1) "Quality Customer Service." In the second stage, statistical procedures were conducted for the independent variable 2 (X2) "Reduced Cost" using simple, multiple, and sequential regression analyses. In the third stage, similar analyses were conducted for independent variable 3 (X3) "Efficiency and Reliability." These three stages were performed separately because Process Macro 4.2 version does not allow the simultaneous inclusion of three independent variables in the analysis. Therefore, each independent variable was defined and analyzed individually.

**Findings**

Table 6 presents the findings of the simple, multiple, and sequential regression analyses conducted to examine the mediating effect of blockchain technology on cryptocurrency purchase intention. Based on the literature, 21 hypotheses were developed and analyzed. The results revealed that quality customer service had a significant and positive effect on perceived usefulness (b = 0.431, 95% CI [0.3259, 0.5769], t = 8.04, p = 0.000). It was also found that quality customer service accounted for approximately 12% of the variance in perceived usefulness, supporting $H_{1a}$. Furthermore, reduced costs had a significant and positive effect on perceived usefulness (b = 0.475, 95% CI [0.3649, 0.5853], t = 8.4693, p = 0.000). Quality customer service accounted for approximately 13.5% of the variance in perceived usefulness, thus supporting $H_{2a}$. Efficiency and reliability had a significant and positive effect on perceived usefulness (b = 0.624, 95% CI [0.5194, 0.7302], t = 11.6471, p = 0.000), explaining approximately 23% of the variance in perceived usefulness and supporting $H_{3a}$. Quality customer service was found to have a significant and positive effect on perceived ease of use (b = 0.111, 95% CI [0.0371, 0.1854], t = 2.94, p = 0.003), thus supporting $H_{1b}$. However, reduced costs did not have a significant effect on perceived ease of use (b = 0.030, 95% CI [-0.0492, 0.1092], t = 0.74, p = 0.47), rejecting $H_{2b}$. Similarly, efficiency and reliability did not have a significant effect on perceived ease of use (b = 0.650, 95% CI [-0.177, 0.1516], t = 1.55, p = 0.12), thus rejecting $H_{3b}$. Perceived usefulness has a significant and positive effect on perceived ease of use (b = 0.643, 95% CI [0.5829, 0.7033], t = 20.94, p = 0.000), supporting $H_6$. Quality customer service and perceived usefulness together accounted for approximately 55% of the variance in perceived ease of use.

Regarding cryptocurrency purchase intention, it was found that quality customer service had a significant and positive effect (b = 0.181, 95% CI [0.0910, 0.2727], t = 3.93, p = 0.001), supporting $H_{1c}$. However, reduced costs did not have a significant effect (b = 0.004, 95% CI [-0.815, 0.911], t = 0.10, p = 0.91), rejecting $H_{2c}$. Similarly, efficiency and reliability did not have a significant effect (b = 0.067, 95% CI [-0.1603, 0.0245], t = -1.44, p = 0.14), thus rejecting $H_{3c}$. Perceived usefulness has a significant and positive effect (b = 0.388, 95% CI [0.2946, 0.4829], t = 8.1, p = 0.000), supporting $H_4$. Perceived ease of use had a significant and positive effect (b = 0.099, 95% CI [-0.0007, 0.1981], t = 1.95, p = 0.05), supporting $H_5$.

For the multiple regression (simple mediation) analysis, six hypotheses were developed, Of all, four were supported but two were rejected. In $H_{7a}$, perceived usefulness was found to mediate the relationship between perceived ease of use and purchase intention (p = 0.001). The interpretation of the analysis results in terms of mediation effects follows the general guidelines, where a fully standardized effect size ($K^2$) close to 0.01 indicates a small effect, $K^2$ close to 0.09 indicates a medium effect, and $K^2$ close to 0.25 indicates a large effect (Gürbüz, 2019: 72). In $H_{7a}$, the value of $K^2 = 0.17$ suggests a medium-level effect. In $H_{7b}$, the analysis revealed a high-level effect, with $K^2 = 0.1847$, indicating that perceived usefulness mediates the relationship between reduced cost and purchase intention. In $H_{7c}$, the analysis showed a high-level effect, with $K^2 = 0.2569$, indicating that perceived usefulness mediates the relationship between efficiency, reliability, and purchase intention. $H_{7d}$ examines the mediating effect of perceived ease of use on the relationship between customer service quality and purchase intention. So, hypotheses $H_{7a}$, $H_{7b}$, $H_{7c}$, $H_{7d}$ are supported. However, there was no effect of perceived ease of use on the effect of reduced costs, efficiency and security on purchase intention. On the other hand, $H_{8a}$, $H_{8b}$, and $H_{8c}$ are supported by sequential mediation analyses, indicating the mediating effects of the variables.

Table 6. Findings Regarding Simple Regression Analysis

| Process Macro: Model 6 | | | | | | |
|---|---|---|---|---|---|---|
| X$_1$:QCS X$_2$:RC X$_3$:ER M$_1$:PU M$_2$: PEU Y: PI | | | | | | |
| Hypothesis | Relationships | B | Bootstrap Confidence Interval Values | p | t | Results |
| H$_{1a}$ | QCS→PU | .431*** | [.3259, .5769] | .000 | 8.04 | Supported |
| | R²= .1228 | | | | | |
| | F (1; 461)= 64,5391; p<.001 | | | | | |
| H$_{2a}$ | RC→ PU | .4751*** | [.3649, .5853] | .000 | 8.4693 | Supported |
| | R²= .1346 | | | | | |
| | F (1; 461)= 71,7292; p<.001 | | | | | |
| H$_{3a}$ | ER→PU | .6248*** | [.5194, .7302] | .000 | 11.6471 | Supported |
| | Efficiency and reliability explain approximately 23% of the variation in perceived usefulness. | | | | | |
| | R²= .2274 | | | | | |
| | F (1; 461)= 135.6561; p<.000 | | | | | |
| H$_{1b}$ | QCS→ PEU | .1112* | [.0371, .1854] | .003 | 2.9484 | Supported |
| H$_{2b}$ | RC→ PEU | .0300 | [-.0492, .1092] | .475 | .7443 | Not Supported |
| H$_{3b}$ | ER→ PEU | .6504 | [-0177, .1516] | .120 | 1.5550 | Not Supported |
| H$_6$ | PU → PEU | .6431*** | [.5829, .7033] | .000 | 20.9889 | Supported |
| | Quality customer service and perceived usefulness together explain approximately 55% of the variation in perceived ease of use. | | | | | |
| | R²= .5497 | | | | | |
| | F (2; 460)= 280,7807; p<.0001 | | | | | |
| H$_{1c}$ | QCS → PI | .1816* | [.0910, .2727] | .001 | 3.9364 | Not Supported |
| | Quality customer service explains approximately 3.25% of the variation in purchase intention. | | | | | |
| | R²= .0325 | | | | | |
| | F (1; 461)= 15.4950; p<.05 | | | | | |
| H$_{2c}$ | RC → PI | .0048 | [-.0815, .0911] | .913 | .1092 | Not Supported |
| H$_{3c}$ | ER → PI | -.0679 | [-.1603, .0245] | .149 | -1,4437 | Not Supported |
| H$_4$ | PU → PI | .3888*** | [.2946, .4829] | .000 | 8.115 | Supported |
| H$_5$ | PEU → PI | .0991* | [-0007, .1981] | .050 | 1.9504 | Supported |
| H$_{7a}$ | QCS → PU → PI | .1692** | [.1032, .2413] | .001 | | Supported |
| H$_{7b}$ | RC → PU → PI | .1847*** | [.1148, .2682] | .000 | | Supported |
| H$_{7c}$ | ER → PU → PI | .2569*** | [.1726, .3502] | .000 | | Supported |
| H$_{7d}$ | QCS →PEU→ PI | .0115** | [-.0036, .0294] | .001 | | Supported |
| H$_{7e}$ | RC → PEU→ PI | .1432 | [-.0089, .0147] | .117 | | Not Supported |
| H$_{7f}$ | ER → PEU → PI | .2395 | [-.0045, .0206] | .185 | | Not Supported |
| H$_{8a}$ | QCS→PU→PEU→PI | .0289** | [-.0072, .0706] | .001 | | Supported |
| H$_{8b}$ | RC→PU→PEU→PI | .0314*** | [-.0092, .0779] | .000 | | Supported |
| H$_{8c}$ | ER→PU→PEU→PI | .0396*** | [-.0037, .0915] | .000 | | Supported |

*p<.05, **p<.01, ***p<.001

## Discussion

This study examines the factors influencing the purchase intention of cryptocurrencies from the TAM perspective, which is widely recognized as one of the most effective models for explaining the adoption of new technologies. Based on the study's findings, the hypothesis that quality customer service has a positive and significant impact on perceived usefulness, perceived ease of use, and purchase intention was supported. These results indicate that service quality has a positive influence on perceived usefulness, perceived ease of use, and purchase intention, consistent with previous studies (Akgül, 2018; Abu-Taieh et al., 2022).

The expense associated with technology assumes a pivotal role in users' discerned worth and the rate of adoption (Premkumar et al., 1997). The financial aspect significantly influences the incorporation of novel advancements by consumers, and past studies have illustrated a clear and significant correlation between cost and the adoption of technology (Seyal et al., 2006). In a similar vein, this investigation unveiled that reduced expenses bestow a positive and notable impact on perceived value. However, the correlation between decreased costs and ease of use or the intention to purchase is not found to be positive. The sequential mediating effect of perceived usefulness showed that reduced costs on blockchain technology had a positive and significant effect on the intention to purchase cryptocurrencies, but not on perceived ease of use. It came to light that factors like lowered transaction, administrative, and operational expenses, heightened transaction efficiency, and enhanced security do not considerably impact purchase intention, except when modulated by perceived value and ease of use. This inquiry makes a scholarly contribution by establishing a connection between cost reduction and the intent to purchase, mediated through the roles of perceived value and ease of use.

The TAM framework has been extensively utilized to explore the reception of diverse technological novelties, with the consistent finding of positive repercussions of perceived value and ease of use on behavioural intent. In the realm of blockchain-based cryptocurrencies, akin to this study, prior research has disclosed that perceived ease of use and perceived value exert a considerable positive influence on behavioural intent (Kamble et al., 2019; Jin et al., 2020; Bharadwaj and Deka, 2021).

An additional notable outcome of this examination is the significant and affirmative sway of perceived value on perceived ease of use. The amalgamation of customer service quality and perceived value accounted for around 5% of the variance in perceived ease of use. In the academic realm, perceived value is acknowledged as a critical determinant impacting the inclination to employ virtual currencies as a payment method, particularly in the domain of cryptocurrencies (Johar et al., 2021).

## Theoretical Contributions

Within the scope of this research, this study examined whether external factors such as quality customer service, reduced cost, efficiency, and security have a positive and significant impact on perceived usefulness, and purchase intention in the context of blockchain technology. The findings of this study indicate that these factors have a positive and significant influence on consumers in Turkey, thus supporting the existing literature.

The primary theoretical contribution of this study is that it provides a foundation for future research focusing on the purchase intention of cryptocurrencies within the domain of blockchain technology. Thus, this study aims to enrich the literature and contribute to the body of knowledge on blockchain technology. While there are several international studies on blockchain applications covering areas such as blockchain technology itself (Nair and Sebastion, 2017), security issues (Lin and Liao, 2017), and challenges and opportunities (Zheng et al., 2017), there are a limited number of studies specifically examining consumer purchase behavior in relation to blockchain technology. Some of these studies briefly analyzed cryptocurrency systems (Mukhopadhyay et al., 2016), Bitcoin as a cryptocurrency (Crosby et al., 2016), and cryptocurrency in financial markets (Legotin et al., 2018).

As digital transformation continues to sweep through every aspect of society, individuals and communities are being compelled to adapt. In particular, there is growing interest in blockchain technology and cryptocurrencies based on blockchain technology. Applications built on blockchain technology can be cost-effective and can help meet regulatory requirements. Moreover, owing to its potential for transparency and traceability, blockchain technology can enhance the efficiency of cryptocurrency purchasing processes (Garg et al., 2021). Therefore, it is crucial to understand the perceptions of professionals and consumers regarding technology and its benefits to develop and implement blockchain technology further (Garg et al., 2021). Traditional database technologies often present various financial and legal obstacles, leading to an increased need for third-party oversight. Hence, this study examines the relationships between external factors (quality customer service, reduced cost, efficiency, and security) and blockchain systems to analyze the benefits of blockchain applications and establish their mediating effects. This contribution is significant for both the literature and industry practitioners.

**Practical Implications**

The outcomes of this investigation offer valuable perspectives for experts within the blockchain and cryptocurrency sectors. A benefit of incorporating blockchain into business procedures is its capacity to streamline operations by lessening unwarranted obstacles among involved parties. Blockchain empowers establishments to foster faith and clarity in business dealings, given that these transactions rely on a digital data recording system as opposed to being overseen by a visible central entity (Garg et al., 2021). Transactions conducted using blockchain technology offer significant benefits to consumers in terms of security, efficiency, and auditability (Crosby et al. 2016). The blockchain technology is a trusted, transparent, and secure system for recording and managing data across various applications. This technology is expected to become increasingly popular for innovative use beyond cryptocurrencies.

Although blockchain technologies and digital currencies are still relatively new, they have gained general acceptance owing to the opportunities they offer. Cryptocurrencies, designed to be used over the Internet, provide businesses and consumers with economic freedom and have a wide range of investment opportunities. Cryptocurrency trading, which is independent of a specific country or centralized institution, offers fast, instant, and independent transactions (World Bank Group, 2018). Cryptocurrencies are virtual and non-physical, and their buying and selling transactions are conducted through digital platforms and recorded on a digital ledger system.

Based on the research findings, businesses and consumers looking to participate in cryptocurrency transactions using blockchain infrastructure should factor in aspects such as quality of customer service, reduced cost, efficiency, and security. Specifically, by concentrating on financial and commercial activities within a blockchain environment that offers these characteristics, consumers can access the desired services expediently, productively, and securely. Emphasizing the advantages of blockchain infrastructure for individuals aged 18 and above, with a strong intent to purchase cryptocurrency, can be beneficial. Additionally, confirming the essential elements of TAM in the context of cryptocurrency trading is worth noting.

From these findings, it is important for businesses and consumers seeking to increase the acceptance level of cryptocurrency purchase intentions based on the blockchain infrastructure to make informed decisions with a certain level of awareness. The cryptocurrency industry facilitated by blockchain technology is widespread globally, and the popularity of cryptocurrency markets, including Bitcoin and Ethereum, is increasing in Turkey. Therefore, prioritizing cryptocurrency products that enhance business and quality of life and improve overall performance rather than relying on services offered by traditional database technologies is crucial. Additionally, designing systems that facilitate easy learning and adoption of cryptocurrencies in our society, as well as highlighting these aspects in marketing activities, would be beneficial.

**Conclusion**

The central aim of this research was to assess the perceived value and simplicity of employing blockchain technology in consumers' intentions to purchase cryptocurrencies, seen through the lens of the TAM framework. TAM has been broadly employed to scrutinize the incorporation of technology by consumers, given its delineation of how individuals adopt and initiate the usage of an innovation (Davis, 1989; Zaineldeen et al., 2020; Amadu et al., 2018). As per Davis (1989), perceived simplicity and perceived value stand as the pivotal factors employed to gauge the acceptance of a system. Collectively, these factors shape attitudes and conduct. In this study, the correlations between external variables, such as exceptional customer service, cost reduction, efficiency, and security, and their influence on pertinent variables and purchase intentions were analyzed.

Based on the demographic findings obtained in this study, it was concluded that male consumers had a slightly higher intention to purchase cryptocurrencies through blockchain technology than female consumers. This indicates that males tend to have more financial wealth and engage in investment activities more than females. Additionally, unmarried individuals were more inclined to purchase cryptocurrencies than married individuals. This suggests that the family concept directs individuals toward investments that are less innovative, risky, and safer. The study revealed that the intention to purchase cryptocurrencies was higher among young adults, particularly those between the ages of 25-34 and 35-44, as well as young individuals between 18-24. This can be attributed to the younger generation being more knowledgeable about the subject and more open to the financial opportunities offered by the digital realm. Moreover, consumers working in the private sector with higher educational and income levels have a higher intention to purchase cryptocurrencies through blockchain technology.

In the analysis of the impact of blockchain technology on the intention to purchase cryptocurrencies, the research findings indicated that quality customer service has a significant and positive effect on the intention to purchase. As Taskin (2015) defines it, customer service is about "meeting customer needs and providing them with service beyond their expectations." Quality customer service involves anticipating customer needs and meeting them promptly when a customer requests a product or service. Within the scope of This study revealed that better and more effective service delivery plays a significant role in the intention to purchase cryptocurrencies. Virtual currencies, as they are not controlled by any central authority, are open to development, change, and provision of better customer service. Therefore, it can be said that new consumer demands that traditional payment methods and financial services struggle to meet can be addressed more quickly and easily with virtual currency (Özkul & Baş, 2020; Yılmaz & Tümtürk, 2015). However, the research findings showed that reduced costs, efficiency, and security, as external factors, did not have a significant and positive impact on the intention to purchase cryptocurrencies. In other words, the visible effects of reduced transaction, administrative, and operational costs as well as efficiency and security concerns related to transactions did not have a significant impact on purchase intention.

Within the TAM framework, the research findings confirmed that perceived usefulness and ease of use had a significant and positive impact on the intention to purchase cryptocurrencies. Both determinants were found to have a positive influence on attitudes and behaviors towards technological systems, acceptance, and usage intentions. As many studies have emphasized, intention to use is primarily determined by perceived usefulness, whereas perceived ease of use is a secondary determinant of intention. It has been widely noted that only these two variables have a direct impact on behavioral intentions. Davis (1989) stated a positive relationship between perceived ease of use and perceived usefulness. Therefore, it can be said that there is a cause-effect relationship between perceived ease of use and perceived usefulness, as individuals who easily learn a specific technological system experience an increase in their job performance. The research findings suggest that perceived ease of use plays a mediating role in the impact of perceived usefulness on the intention to purchase. In other words, the research findings indicate that perceived ease of use is an important factor influencing individual intentions in the usage process of technological systems. This research highlights the strong influence of perceived ease of use on the structure of perceived usefulness and the significant impact of quality customer service on perceived ease of use.

Traditional database services are often costly to access in today's rapidly advancing technological landscapes. Cryptocurrencies enable faster and cheaper transactions (Özkul & Baş, 2020). Perceived ease of use has a direct impact on the use of information technologies (Chiu et al., 2009). This research demonstrates that the perceived usefulness of blockchain technology and the ease of understanding and using cryptocurrencies have a mediating effect on their efficiency and security. Consequently, the security concerns associated with cryptocurrency transactions and the efficiency achieved through various security measures and innovative security layers mediate the intention to purchase.

The research findings indicate that quality customer service significantly and positively influences the intention to purchase cryptocurrencies through its mediating effect on perceived ease of use. Therefore, the research suggests that perceived ease of use plays a crucial role in the relationship between quality customer service and the intention to purchase. It can be concluded that perceived ease of use is an important factor in shaping individual intentions in the usage process of technological systems. The research highlights the strong influence of quality customer service on the structure of perceived ease of use and its significant impact on the intention to purchase.

In conclusion, the research findings demonstrate the critical role of perceived usefulness and perceived ease of use in shoppers' intention to purchase cryptocurrencies through blockchain technology. This study furnishes valuable insights into the influences that affect customers' utilization of blockchain technology and sheds light on the mediating effects of perceived ease of use. It is important for businesses and consumers seeking to enhance their cryptocurrency purchase intentions to consider the factors of quality of customer service, perceived ease of use, and perceived usefulness. These findings contribute to the existing literature on consumers' intention to use cryptocurrencies and provide practical implications for businesses and individuals aiming to adopt blockchain technology for cryptocurrency transactions.

Regarding the limitations of the study, it should be noted that the data were collected through convenience sampling from individuals aged 18 and above residing in different regions of Turkey. To enhance the generalizability of the results, future studies could employ random sampling methods. Additionally, the study focused on individuals interested in cryptocurrency through online platforms, and questions related to quality customer service, reduced costs, efficiency, and security were directed towards these individuals. Future research could delve more deeply into specific cryptocurrencies or specific online platforms. It would also be beneficial to examine consumers' purchase intentions and attitudes based on demographic differences, such as specific occupational groups or individuals with certain income levels, to obtain more detailed findings and contribute positively to the literature.

**Limitations and Future Research**

This study has several limitations. Due to limited research capacity, time, and energy, the research was conducted online. Additionally, the sample of the study was limited to Turkey only. Conducting a similar study in different countries would expand the literature. Furthermore, considering the sample size in this study, future research could make the study more comprehensive by including a larger number of participants and collecting data over a longer period. Blockchain and cryptocurrencies are vast research areas, and the factors influencing purchasing and the pathways of influence are complex. This study represents only a fraction of the factors. Conducting future research while considering different factors would strengthen the literature. Therefore, future research should focus on creating a more comprehensive model of cryptocurrency purchase intention and exploring other boundary conditions to enhance the validity of the model.

The positive factors arising from the free open source structure in the nature of blockchain-based cryptocurrencies along with features such as transaction control and high efficiency enhance the perceived usefulness and perceived ease of use. By examining the effects of quality customer service, reduced cost, efficiency and reliability factors within the framework of the TAM, our results open up an exciting new field of study. Other antecedent factors can be added to the study in future studies.